\documentclass[10pt]{iopart}

\usepackage[T1]{fontenc}
\usepackage[utf8]{inputenc}
\usepackage{float}
\usepackage{cuted}
\usepackage{graphicx}
\usepackage{tabularx}
\usepackage{color, soul}
\usepackage{textcomp}
\usepackage{enumerate}
\usepackage{todonotes}
\usepackage[draft]{hyperref}
\usepackage{comment}
\usepackage{subcaption}
\usepackage{url}
\usepackage[normalem]{ulem}
\usepackage{makecell}
\usepackage{cite}
\usepackage{caption}
\usepackage{subcaption}
\usepackage{pifont}
\usepackage{dblfloatfix}
\usepackage{longtable}
\usepackage{multicol}
\begin{document}
\captionsetup[figure]{labelfont={bf},labelformat={default},labelsep=period,name={Fig.}}

\title[]{Inhibition and promotion of quasi--uniform to filamentary discharge transition in negative repetitive nanosecond surface dielectric barrier discharge}

\author{Yaqi Zhang$^{1}$, Yulin Guo$^{1}$, Yifei Zhu$^{1,2}$, Anbang Sun$^1$}

\address{$^1$ State Key Laboratory of Electrical Insulation and Power Equipment, School of Electrical Engineering, Xi'an Jiaotong University, Xi'an, 710049, China\\
	$^2$  Institute of Aero-engine, School of Mechanical Engineering, Xi’an Jiaotong University, Xi’an 710049, People's Republic of China}

\ead{yifei.zhu.plasma@gmail.com; anbang.sun@xjtu.edu.cn}

\vspace{10pt}
\begin{indented}
        \item[]\today
\end{indented}

\begin{abstract}
The transition from quasi--uniform to filamentary modes in a repetitive nanosecond Surface Dielectric Barrier Discharge(SDBD) under atmospheric pressure was studied. Our focus encompassed both discharge morphology and electrical characteristics, revealing two pivotal findings. Firstly, by analyzing the current and the deposited energy waveforms, three characteristic frequency ranges respectively corresponding to the discharge modes are identified. Notably, within the 5~kHz to 8~kHz range, we observed non-monotonic changes in the propagation distance, the current amplitude, and the deposited energy---a crucial insight linked to the discharge transition. Secondly, the count of current extrema in the primary discharge process changes only during the transition to filamentary mode, remaining stable in the steady discharge mode. This variation may be attributed to secondary Surface Ionization Waves (SIWs). The interplay of these two findings with the discharge transition requires deeper investigation. Additionally, we present a discharge modes control curve outlining parameter windows for the discharge modes. This curve facilitates the optimization of pulse power supply and control schemes in practical applications.
\end{abstract}

\ioptwocol

% 1. Introduction
% -  The applications of the SDBD
% -  
\section{Introduction}
\label{sec:introduction}
Surface dielectric barrier discharge (SDBD) has garnered significant attention in various fields, such as ignition and combustion~\cite{starikovskaia2014plasma,ju2015plasma,yin2011ignition,zhao2022experimental}, active flow control~\cite{leonov2016dynamics,adamovich2012nanosecond,benard2014electrical}, sterilization and disinfection~\cite{ambrico2020surface,mitsugi2018practical,daeschlein2012skin}, and surface material treatment~\cite{borcia2003dielectric,niu2013repetitive}. The uniformity of SDBD is a significant concern: in some industries, planar uniformity is required to achieve large--scale treatment, while filamentary discharges are preferred for higher energy density in other applications.

Nanosecond pulse generators have been used to drive SDBDs to achieve planar uniformity in many groups. Existing researches~\cite{Walsh2007Room,walsh200710ns,mizuno1984device,zhang2016needle} had highlighted the advantageous characteristics of single--pulse nanosecond surface dielectric barrier discharge (SDBD). With its short pulse width, rapid rise, and highly reduced electric field, this technique predominantly channeled energy into high--energy electrons. This unique feature enhanced the efficiency of chemical processes like ionization and excitation in gases. Meanwhile, the short pulse duration minimizes temperature increase, avoiding issues like severe thermal effects common in repetitive pulsed discharge. The single--pulse discharge can easily generate large--area and uniform plasma, making it a compelling choice. 

However, with the development of high frequency, high amplitude pulse generators, researchers have identified that under specific conditions, quasi--uniform to filamentary transition could occur even in nanosecond SDBDs~\cite{stepanyan2014nanosecond,ding2019filamentary,soloviev2022filament}.

% Why does nanosecond SDBD have high uniformity in one pulse? 

One of the typical conditions of the transition was found to be high pressure and high voltage amplitude. The transition occurs in the first pulse. Stepanyan et al.\cite{stepanyan2014nanosecond} and Ding et al.\cite{ding2019filamentary} explored the quasi--uniform filament transition process, investigating various factors such as N$_2$:O$_2$ mixing ratios, air pressures (1--12~bar), voltage amplitudes (20--55~kV), and voltage polarities. They provided voltage and pressure coordinate transition curves, revealing that a lower oxygen ratio, higher air pressure, and increased voltage amplitude facilitate the transition. The threshold voltage for transition decreases with rising pressure. Stepanyan~\cite{stepanyan2014nanosecond} proposed ionization heating instability at the cathode layer boundary as the mechanism for filament formation. 

Stepwise ionization and dissociation of excited states of molecules are crucial processes that drive the transition from streamer to filament in nanosecond discharges. Soloviev~\cite{soloviev2022filament} employed a two-dimensional fluid model to simulate the development of the SDBD in an N$_2$ atmosphere with 0.1~\% O$_2$ at pressures ranging from 2 to 8~bar and voltages between 20 to 50~kV. Increasing air pressure and voltage amplitude resulted in the formation of flow and dielectric boundaries, which created a thin layer of high electron density, believed to represent the filaments. The pressure--voltage transition curve obtained from simulations was consistent with experimental observations, attributing the nanosecond SDBD transition under positive polarity to stepwise ionization of nitrogen molecules in excited states. Similar conclusions have been drawn for air as well~\cite{zhang2023streamer}.

Another condition triggering the transition was atmospheric pressure but high pulse frequency. The transition occurs after a series of pulses. Takashima et al.\cite{takashima2011characterization} and Pang et al.\cite{lei2016discharge} observed a noteworthy phenomenon: the transition of SDBD at lower voltage amplitudes (\textless 20~kV) under repeated nanosecond pulses at atmospheric pressure. As the pulse number or repetition frequency rose, the discharge transformed from the uniform to the filamentary mode. Experimental and theoretical studies by Huang et al.\cite{huang2014influence,huang2018effect} emphasized the impact of residual charges on breakdown voltage reduction and discharge enhancement in repetitive pulse scenarios. Naidis~\cite{naidis2008simulation} simulated atmospheric pressure with a 5~kV voltage amplitude and a repetition frequency ranging from 15~kHz to 30~kHz. With increased pulse frequency or number, the discharge modes transitioned from diffuse to spark.

While previous studies~\cite{nijdam2014investigation,li2018positive,ren2023impact,kumada2009residual,popov2021repetitively,tholin2013simulation,janda2017influence,ndong2013effect,zhao2020volume} have explored the impact of repetition frequency and pulse number, these investigations often feature a limited frequency range and predominantly offer qualitative analyses rather than quantitative insights. To address this gap and provide comprehensive guidance for diverse applications, our research extends into a broader frequency spectrum, aiming to deliver detailed quantitative and qualitative results. We focus on unraveling specific phenomena associated with transitions, offering a thorough description and explanation of the underlying processes. This approach aims to enhance the applicability of the findings and contribute valuable insights for a wider array of applications.

This study aims to quantify the influence of the repetition frequency and the number of pulses on the electrical characteristics, the discharge morphology, and the evolution of nanosecond SDBDs under atmospheric pressure. It helps to understand the mechanisms of quasi--uniform to filamentary discharge transition of nanosecond SDBDs for flexible design and control SDBD devices. Get the Number--Frequency(N--F) transition curve as a refined control method to enhance the effectiveness of applications.

The structure of this paper is as follows: Section 2 unveils the experimental setup; Section 3 presents the experimental results encompassing discharge morphology, evolution processes, electrical characteristics, pulse parameters(repetition frequency and pulse number) effects, and distinctions between the two main discharge modes: quasi--uniform and filamentary discharge. Finally, the N--F transition curve is presented. %The Number--Frequency(N--F) curves serve as a pivotal tool for precisely regulating the transition of discharge modes or maintaining stability in various applications. This refined control method can enhance the effectiveness of applications.

\section{Experimental and Numerical Methods}
\label{sec:Setup}

\begin{figure}[H]
\renewcommand{\figurename}{Fig.}
\centering
\includegraphics[width=83mm]{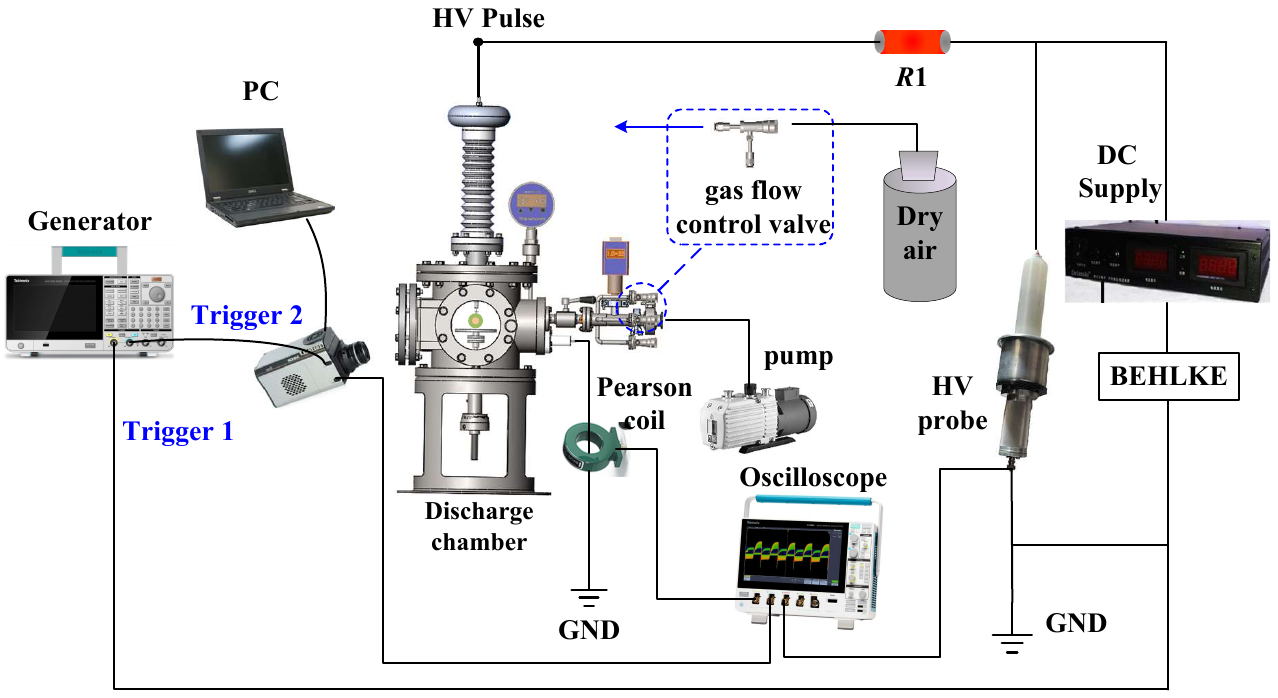}
\caption{Scheme of the experimental setup}
\label{fig:setup}
\end{figure}

A systematic measurement of the pulse repetition frequency and current in negative repetitive SDBD has been conducted in this work, aiming to lock the key parameters affecting the transition. The deposited energy, the number of pulses, and the morphology when the transition occurs were tracked. The experimental platform, illustrated in Fig.~\ref{fig:setup}, comprises a gas control system, a pulse plasma generation system, an image acquisition system, and a measurement system. 

The gas control system is used to fill the discharge chamber with dry air. The chamber is evacuated to below $10^{-3}$~Torr and then pressurized to 1~bar before each experiment. 

\begin{figure}[H]
\renewcommand{\figurename}{Fig.}
\centering
\includegraphics[width=83mm]{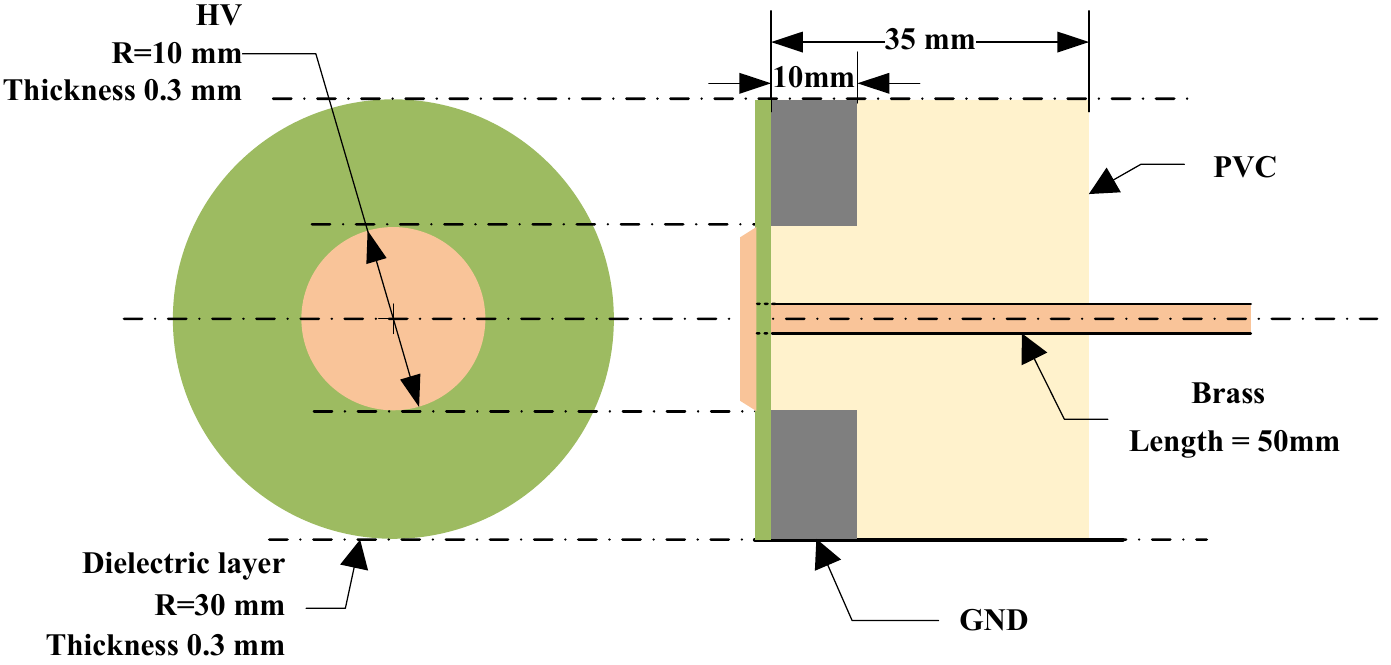}
\caption{Geometry of the SDBD electrode system}
\label{fig2:nSDBD}
\end{figure}

The pulse plasma generation system includes a power supply and a coaxial SDBD electrode system. The power supply consists of a high--voltage DC power supply (R2--P60W1k), a 70~kV energy storage module, and a BEHLKE switch module (BEHLKE, HTS 901--10--GM). It incorporates an external current limiting resistor R1 (200~$\Omega$) and is triggered by channel 1 of the signal generator (Tektronix AFG31000). Adjusting the output voltage of the high--voltage DC power supply generates a pulse voltage with an amplitude of --50~kV--0 and the pulse width (5~$\%$ --95~$\%$) of about 400~ns. The coaxial electrode structure is depicted in Fig.~\ref{fig2:nSDBD}, consisting of a thin brass disc as the HV electrode, a dielectric layer (PTFE, $\varepsilon$ = 4), and an alumina hollow cylinder as the ground electrode. Surface discharge emission images are captured and transmitted to a portable laptop using an Intensified Charged Coupled Device (ICCD, Andor iStar DH334T--18U--A3). The ICCD is triggered by signal generator channel 2, with channels 1 and 2 set as synchronous signals.

The measurement system comprises an electrical measurement part and an optical measurement part.  The applied voltage waveform is measured using a high--voltage probe (North Star PVM--4), while the current is measured through a Pearson coil (response time $\le$ 20~ns). Synchronous processing of the measurement and image acquisition systems is achieved using a digital oscilloscope (Tektronix 3 Series MDO34) with a bandwidth of 1~GHz and a maximum sampling rate of 5~GS/s. The oscilloscope records and stores the high--voltage probe voltage, current, and ICCD trigger signal.

% 3 Results and Discussion

\section{Results and Discussion}
% content summary
The study explored the impact of adjusting the repetition frequency(25~Hz to 30~kHz) and the number of pulses(2 to 20) on the morphological features, employing the emission images. With the fixed voltage amplitude(--16~kV), through a blend of qualitative and quantitative analyses, we identified three distinct discharge modes: quasi--uniform, transitional, and filamentary. Our exploration of the electrical characteristics, including the current and the deposited energy, unveiled three distinct evolving trends that precisely aligned with three unique discharge modes.

% criteria of filament formation
To distinguish the discharge modes, we define the following criteria of filament formation: 1) filament spacing should be more than 1~mm; 2) filament diameter should be larger than 0.3~mm; 3) average current value should exceed 0.6~A.

% 3.1 Discharge Evolution and Morphology
\subsection{Discharge Evolution and Morphology}
\begin{figure*}[htb]
\renewcommand{\figurename}{Fig.}
\centering
\includegraphics[width=170mm]{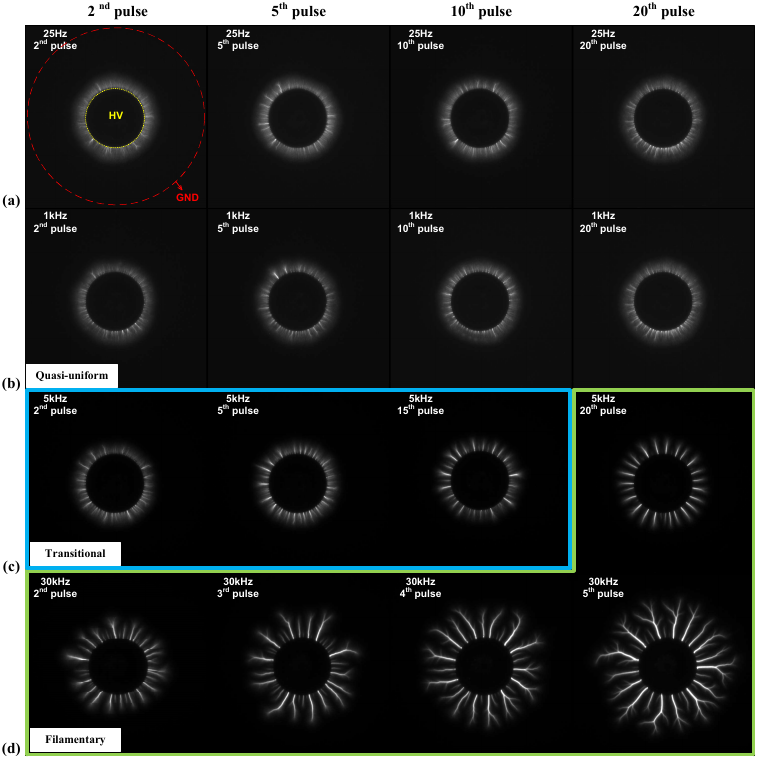}
\caption{The discharge emission images under different pulse repetition frequencies and pulse numbers, each image is the cumulation of the last pulse after applying a burst pulse train.\\
(a) 25~Hz, (b) 1~kHz, (c) 5~kHz, (d) 30~kHz, ICCD exposure time is 450~ns}
\label{fig3:fandN}
\end{figure*}

The emission images, which vary from the repetition frequencies and the pulse numbers captured by ICCD, are shown in Fig.~\ref{fig3:fandN}. These images, provide an efficient and visually intuitive way to comprehend the details of the discharge evolution and morphology. 

The emission images in Fig.~\ref{fig3:fandN} unveil the evolution and morphology of the nanosecond SDBDs, offering insight into their characteristics at different repetition frequencies((a)25~Hz, (b)1~kHz, (c)5~kHz, (d)30~kHz) and pulse numbers($2^{nd}$ pulse -- column one, $5^{th}$ pulse -- column two, $10^{th}$ pulse -- column three, $20^{th}$ pulse -- column four). 

% mark what the discharge modes they belong to in the Fig3
Emission images identified three distinct discharge modes: quasi--uniform, transitional, and filamentary. In Fig.~\ref{fig3:fandN}, the top two rows showcase quasi--uniform modes (top two rows), transitional modes (within the blue line), and filamentary modes (within the green line).

In the qualitative analysis, the quasi--uniform discharges manifested under specific conditions. Notably, they occurred at high repetition frequencies coupled with few pulse numbers (5~kHz -- < 15 pulses, 30~kHz -- < 3 pulses) or exclusively at low frequencies (25~Hz or 1~kHz). While initially presenting as a uniformly discharged area, closer scrutiny unveiled the presence of micro-streamers with consistent propagation velocities. Intriguingly, even with an escalation in repetition frequency and pulse number, the longest propagation distance remained constant, accompanied by a sustained weak emission intensity.

Increasing the repetition frequencies and pulse numbers triggered the transitional discharge, illustrated in the first three columns of Fig.~\ref{fig3:fandN}(c) and the first column of Fig.~\ref{fig3:fandN}(d). This mode is distinguished by luminous contraction channels amidst the diffuse background emission. Notably, these channels, though elongated and brighter than the quasi--uniform discharge, maintain an overall uniformity in the main emission area.

\begin{figure}[H]
\renewcommand{\figurename}{Fig.}
\centering
\includegraphics[width=85mm]{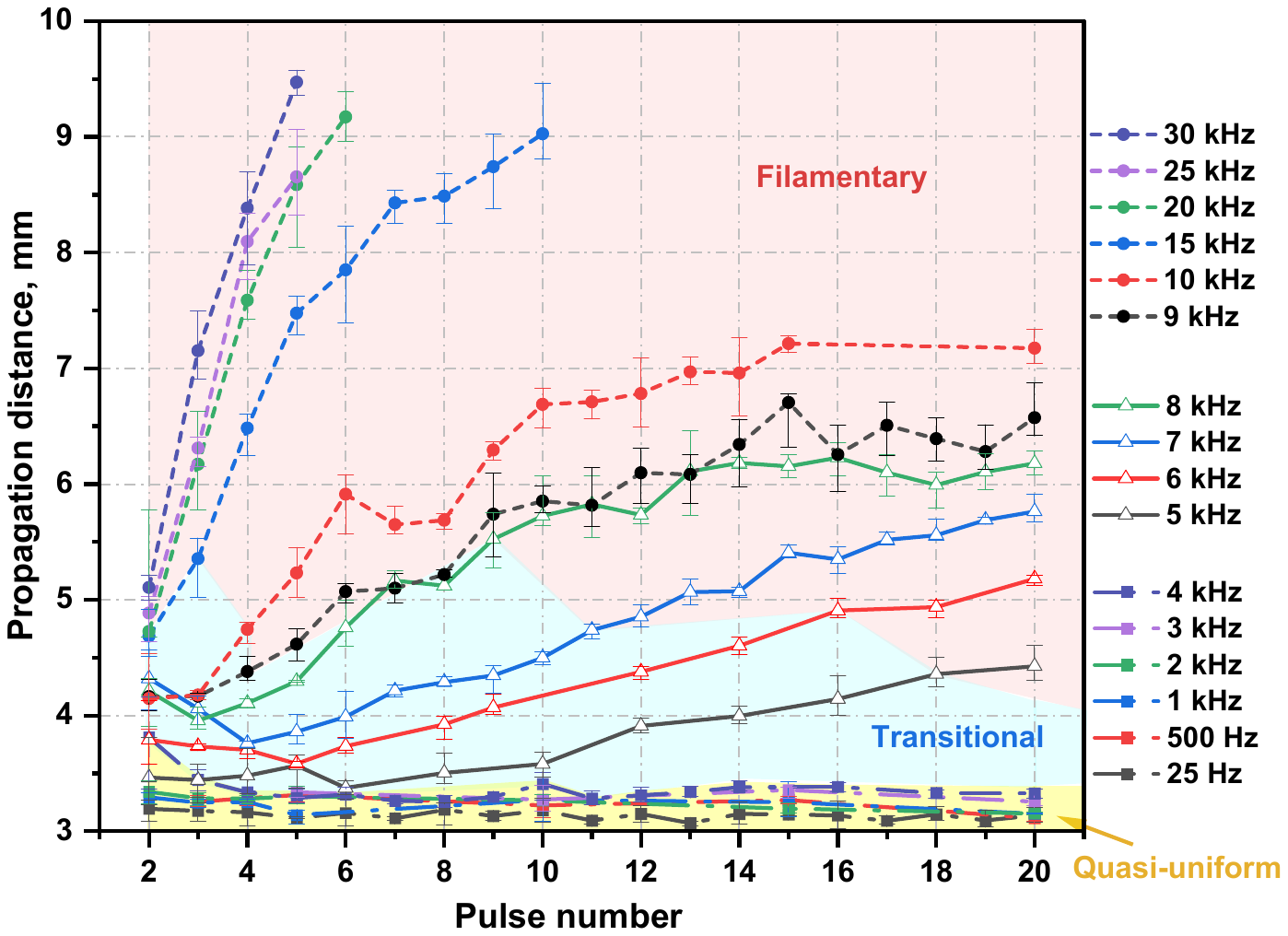}
\caption{The discharge propagation distance from the HV electrode along the dielectric layer varied with the number of pulses and repetition frequency.}
\label{fig4:distance}
\end{figure}

The continued increase of the repetition frequencies and pulse numbers led to the filamentary discharge, depicted in the final column of Fig.~\ref{fig3:fandN}(c) and the first three columns of the Fig.~\ref{fig3:fandN}(d), featuring branching filament heads. These filaments exhibited an emission intensity over three times that of quasi--uniform discharge, with a stable distribution of approximately 1~mm between them. The longest propagation distance extended further with higher repetition frequency and pulse numbers. However, the continuously increased plasma channel eventually connected the HV electrode and the ground electrode, causing surface flashover.

% Quantitative analysis: Propagation distance 
To quantify the aforementioned observations, we conducted a statistical analysis of the images, the relationship among the discharge propagation distance $L$, repetition frequency, and number of pulses was illustrated in Fig.~\ref{fig4:distance}.

% quantitative analysis
% Three discharge modes 
The discharge modes were identified based on acquired emission images. In Fig.\ref{fig4:distance}, the red, blue, and yellow parts represent filamentary, transitional, and quasi--uniform modes, respectively. Notably, with the frequency increased, the transition between discharge modes occurred more efficiently, requiring fewer acquired pulses to occur.

% Overall view: three distinct trends (monotonically decreased, first decreased then increased, monotonically increased) --- the second trend(phenomena) has not been mentioned before.
Three distinct characteristic repetition frequency ranges can be identified by combining Fig.~\ref{fig3:fandN} and Fig.~\ref{fig4:distance}: 

1) From 25~Hz to 4~kHz: When the number of pulses was increased from 2 to 20, $L$ exhibited a minimal decrease, not exceeding 0.47~mm. Meanwhile, as the repetition frequency rose from 25~Hz to 4~kHz, we noticed a slight increase in the average value of $L$ and average velocity per pulse. $L$ rose from 3.15~mm to 3.36~mm, and the velocity increased from $5.3 \times 10^{-2}$~mm/ns to $5.6 \times 10^{-2}$~mm/ns. This repetition frequency range corresponds to the long pulse interval time, leading to most memory factors dissipating. So the memory effect impacted the subsequent discharges little.

2) from 5~kHz to 8~kHz: The propagation distance initially decreased and then increased with the number of pulses. When the pulse repetition frequency increased from 5~kHz to 8~kHz, the average velocity increased from $6.3 \times 10^{-2}$~mm/ns to $9 \times 10^{-2}$~mm/ns. It should be noted that the number of pulses with the smallest discharge propagation distance was inversely proportional to the pulse repetition frequencies (e.g., at 5~kHz--sixth pulse; at 8~kHz--third pulse). The observed phenomenon might can be attributed to the differential growth rates of the effect memory: promotion factors and inhibition factors. With increasing frequency, the promotion factors outpaced the inhibitory ones, leading to a distinctive pattern in the propagation behavior—--initial increased followed by a subsequent decrease.

3) from 9~kHz to 30~kHz: we noticed the average discharge propagation distance gradually increased with both the repetition frequency and the number of pulses. Surface flashover occurred after the frequency exceeded 15~kHz. Specifically, the average discharge propagation velocity increased from $9.4 \times 10^{-2}$~mm/ns to $12.5 \times 10^{-2}$~mm/ns.

%% I want to express that every single line has a different trend with each other, but they increased with the repetition frequencies.  
Noticeably, the average velocity and the propagation distance exhibited a consistent uptrend with the frequency, as depicted in Fig.~\ref{fig4:distance}. However, not each curve monotonically changed, they unfolded their unique pattern within three distinct repetition frequency ranges. In the transitional discharge stage(2 to 7 pulses), a unique phenomenon emerged --— the curve exhibited a non-monotonic behavior, defying conventional expectations. Contrastingly, in the quasi--uniform and filamentary discharge modes, a monotonic evolution of the curve was observed. Thus, we can know that it may relate to the transition of the quasi--uniform to filamentary discharge. So this phenomenon warrants thorough investigation.

Evidently, the average velocity and propagation distance displayed an upward trend with the frequency, as illustrated in Fig.~\ref{fig4:distance}. Interestingly, each curve revealed a distinctive pattern within three specific frequency ranges. These three frequency ranges are aligned with three discharge modes. The special phenomenon observed is the non-monotonic change in the transitional discharge —-- unlike the quasi--uniform and filamentary discharges. This unique behavior hints at a connection to the transition from quasi--uniform to filamentary discharge, prompting the need for in--depth investigation.

% In summary, lower repetition frequencies and fewer pulse numbers inhibit discharge, while higher repetition frequencies and more pulse numbers promote discharge.

\subsection{Comparison of Quasi–uniform Discharge and Filamentary Discharge Characteristics}
% chapter summary

In this section, the quasi--uniform and filamentary discharge modes were compared from two aspects: the morphology(frontal and side view) and the electrical characteristics(current, deposited energy).

\begin{figure}[H]
\renewcommand{\figurename}{Fig.}
\centering
\includegraphics[width=83mm]{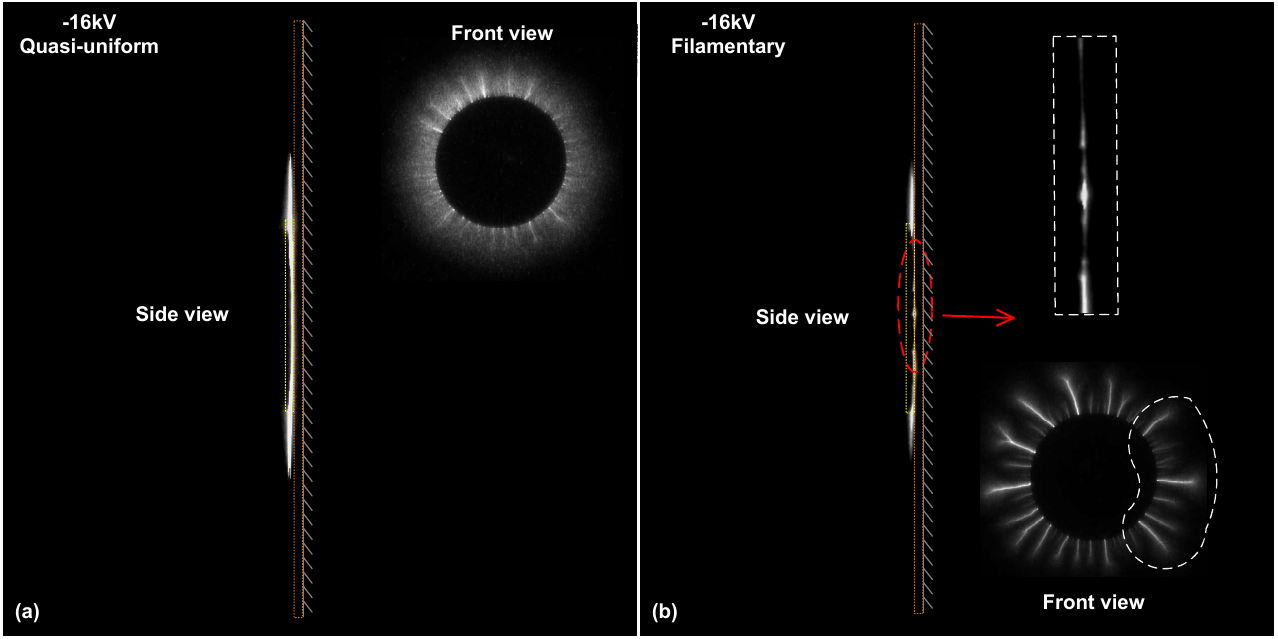}
\caption{Morphology of quasi--uniform discharge and filamentary discharge (The front and the side view). (a) quasi--uniform discharge, (b) filamentary discharge }
\label{fig5:view}
\end{figure}

At first, the morphology of quasi--uniform and filamentary discharges from the front (a) and side (b) is shown in Fig.~\ref{fig5:view}. 

In the frontal view, quasi-uniform discharge had a large, uniform emission area with consistent head propagation velocity; Filamentary discharge displayed varying head propagation velocities, and filaments emitted significantly more than adjacent diffuse areas, with spacing greater than 1~mm. 

In the side view, the plasma layer of quasi--uniform discharge became thinner away from the HV electrode, developing near the surface. Similarly, the plasma layer of filamentary discharge also thinned with distance from the HV electrode. A closer observation of the part intercepted within the red dotted circle revealed the filament propagation close to the surface, with a slightly thinner thickness compared to the quasi--uniform plasma layer. 

\begin{figure}[H]
\renewcommand{\figurename}{Fig.}
\centering
\includegraphics[width=83mm]{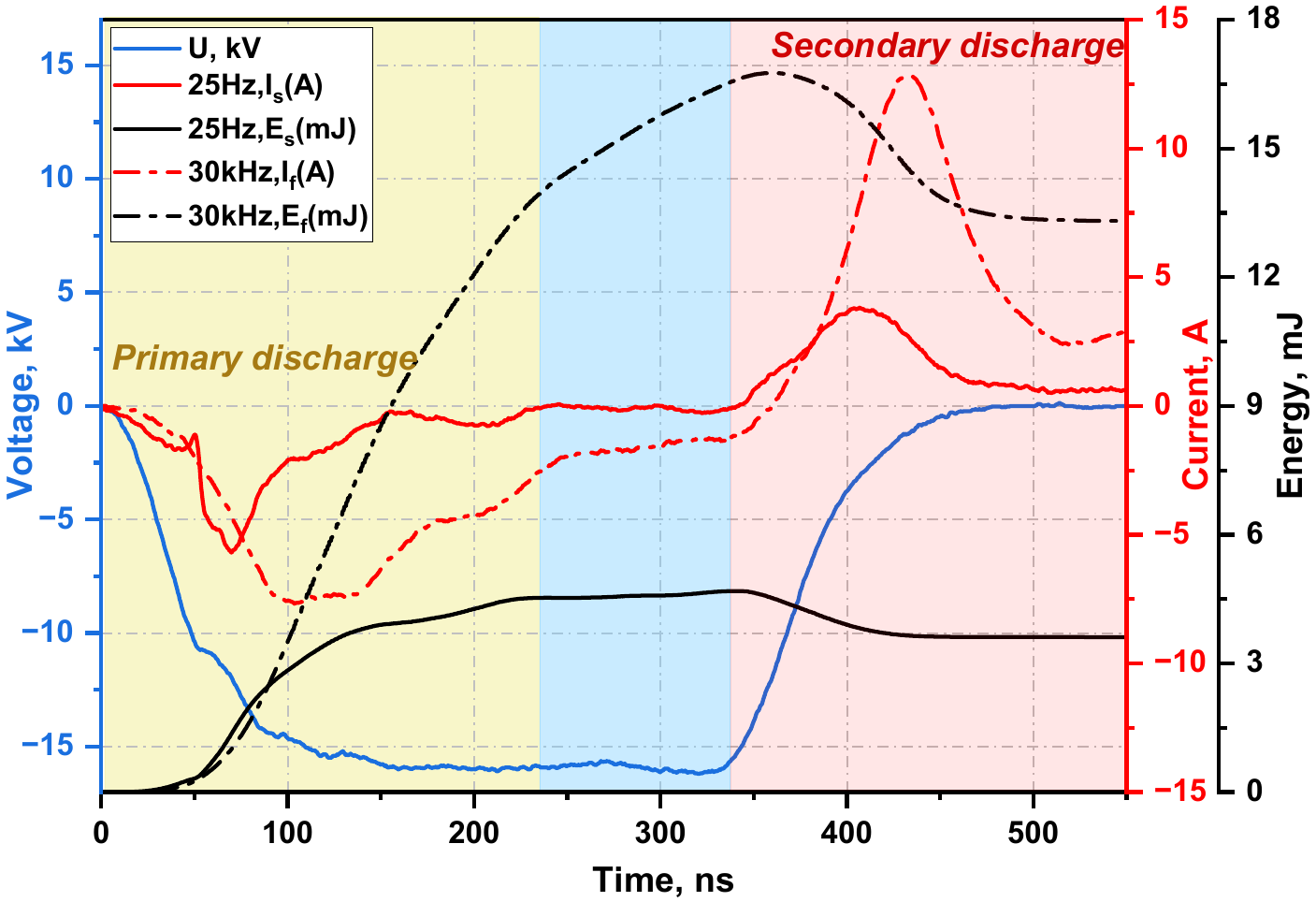}
\caption{Electrical characteristics (voltage, current, deposited energy) in quasi--uniform discharge(25~Hz, solid curve) and filamentary discharge(30~kHz, chain curve).}
\label{fig6:UIE}
\end{figure}

The electrical characteristics corresponding to the two discharge modes in Fig.~\ref{fig5:view} are shown in Fig.~\ref{fig6:UIE}. We presented waveforms of voltage(blue line), current(red line), and deposited energy(black line), and compared the disparities. 

The solid curve signifies the quasi--uniform discharge, while the chain curve denotes filamentary discharge. The waveforms measured at the fourth pulse, at frequencies of 25~Hz and 30~kHz, are selected to represent the typical waveforms in quasi--uniform and filamentary modes. (There is no characteristic requirement for the number of pulses. This work chose the fourth pulse when both discharge modes exist and the morphology is typical.)

% two discharge processes in single pulse: primary and secondary discharge 
Observations indicated that both discharge modes manifested two processes: 1) primary discharge (forward discharge, yellow area): aligned with the applied electric field; 2) secondary discharge (reverse discharge, red area): opposite to the applied electric field.

The inception discharge voltage for both modes was approximately --500~V. In the case of quasi--uniform discharge, the current peaks for both primary and secondary discharges were comparatively lower than those observed in filamentary discharge. Specifically, the current peaks for quasi--uniform discharge were --5.22~A (primary) and 3.81~A (secondary), while for filamentary discharge, they were --7.88~A (primary) and 12.68~A (secondary).

Moreover, the deposited energy exhibited a gradual increase to a stable value in the primary discharge, followed by a decrease in the secondary discharge until stability was achieved. The stable deposited energy for quasi--uniform and filamentary discharge was 3.7~mJ and 10~mJ, respectively.

% prompt the following content
Exploring the disparities in morphology and electrical characteristics between quasi--uniform and filamentary discharges revealed intriguing distinctions. The below investigations will delve into the effect of pulse parameters, including the repetition frequency and the pulse number.

% Effect of the Pulse Parameters
\subsection{Effect of the Pulse Parameters}

Adjusting the pulse parameters is a swift and versatile means to set discharge modes effectively. Therefore, the effects of pulse parameters (repetition frequency and pulse number) on the current and the deposited energy were qualitatively and quantitatively analyzed.

\subsubsection{Effect of Pulse Repetition Frequency}

% Why investigate the repetition frequency?
Variations in pulse repetition frequency directly correlate with the pulse interval time. Within these intervals, the types and concentrations of the memory effect factors undergo dynamic shifts, exerting a crucial effect on the overall transition process.

\begin{figure}[H]
\renewcommand{\figurename}{Fig.}
\centering
\includegraphics[width=83mm]{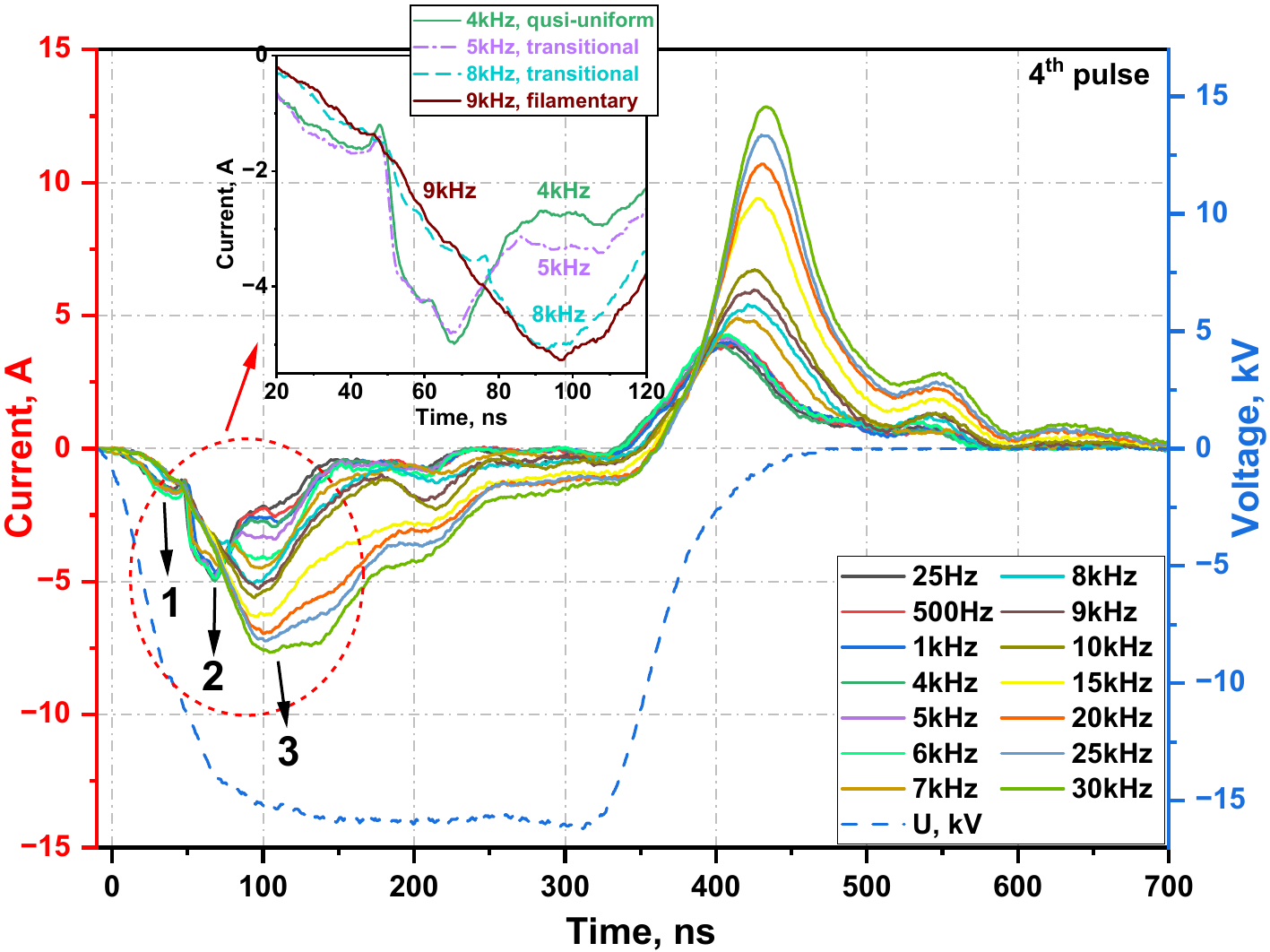}
\caption{Discharge currents across varying repetition frequencies with fixed pulse numbers and voltage amplitude. (The fourth pulse and --16~kV) }
\label{fig7:Current}
\end{figure}

% The trends of the currents at different frequencies
% - for every single line
The effect of the repetition frequency (pulse interval time) on current is shown in Fig.~\ref{fig7:Current}. Each curve in the trend exhibited distinct primary and secondary discharge processes, manifesting during the rising and falling edge of the pulse, respectively.

% Fig7：Current changing with the frequency.

% - for the comparison of all lines
Taking a comprehensive view, we observed the current amplitude increased with the frequency. The time required for primary and secondary currents to return to zero extended noticeably. The evolving patterns of all curves can be categorized into three frequency ranges: 25~Hz to 4~kHz, 5~kHz to 8~kHz, and 9~kHz to 30~kHz, respectively associated with quasi--uniform, transitional, filamentary discharges.

As the discharge mode transited from the quasi--uniform to the filamentary, an intriguing phenomenon emerged in the primary discharge process.

The primary discharge current underwent a distinctive evolution, transitioning from three extrema to a singular peak. In the quasi--uniform mode, the first two extrema remained relatively stable, gradually diminishing during the transitional mode, and ultimately fading in the filamentary mode. Meanwhile, the third extremum exhibited a unique behavior, steadily increasing with frequency.

To illustrate this transformation, refer to the detailed zoom figure depicting the shift from quasi--uniform to filamentary mode (4~kHz to 9~kHz) in Fig.~\ref{fig7:Current}. During the transition from quasi--uniform to transitional mode, only the third extremum registered an increase with frequency. Beyond 8~kHz, the first two extrema decreased, and the third extremum rose. The final shift to the filamentary mode at 9~kHz witnessed the complete disappearance of the first two extrema, leaving only the continuous ascent of the third extremum.

% The cause of the first phenomenon: secondary SIWs. Deny the ionization-thermal instability. 
% two-current-spike phenomenon correlates to the secondary SIW
Note that we observed that the secondary SIWs might be the mechanism of the discharge mode transition. ~\cite{jiang2013experimental} found the difference in the current waveform of two typical discharge modes(diffuse--like mode and multi--streamer,  corresponding to the quasi--uniform mode and the filamentary mode in our experiment). The research showed that the first current extremum was smaller in the diffuse--like mode. However, the first current extremum was higher than the second one in the multi--streamer mode. The continued research~\cite{jiang2021numerical} proved that the two current extrema correspond to the two stages of the discharge: the ionization wave propagation and the repeated re--ignition in the gap between the ionization wave and the dielectric surface using numerical study. 

% FGH is not the mechanism for transition
Zhu et al.~\cite{zhu2020secondary} provides crucial insights by quantifying specific electron density conditions for the emergence of the secondary SIWs. Their study establishes that the secondary SIWs manifest when the electron density remains below $3-5 \times 10^{19}$~m$^{-3}$ during the rising edge of the pulse. This condition arises because the electric field behind the ionization head is insufficient to shield the increasing applied field.

% Additionally, the investigation explores the role of Fast Gas Heating (FGH) in causing the secondary SIWs. FGH is of particular interest due to its potential to induce ionization--thermal instability in some research, a significant process in nanosecond pulsed discharges. Surprisingly, even after conducting calculations with the gas heating module closed in the code—solely solving plasma equations (Poisson’s equation and drift--diffusion--reaction equations) —-- the results confirm that FGH does not cause the secondary SIWs under these conditions. This finding differs from the widely accepted transition mechanism that ionization–-thermal instability drives the discharge mode transition, introducing a perspective to the understanding of the mechanism of the discharge mode transition.

% summary
In our experiment, we considered that the transition in discharge mode is intricately tied to the evolution of current extrema, particularly the third extremum (the secondary SIW or the second current spike, as discussed in ~\cite{jiang2013experimental, jiang2021numerical, zhu2020secondary}). The cause of the secondary SIW stems from the repeated re--ignition in the gap between the ionization wave and the dielectric surface. 

Noticeably, while the primary discharge process in~\cite{jiang2021numerical} reveals two current spikes, our observations present three distinct extrema. This discrepancy raises two potential explanations: 1) Experimental device inaccuracies may account for the disparity in our findings. 2) Alternatively, the amplitude of the first current extremum might have been too subtle to detect in their experiments. We lean towards the former explanation, aligning the last two extrema in our study with the two current spikes documented in~\cite{jiang2021numerical}.

While the secondary SIWs may account for the current extrema in the quasi--uniform discharge, it falls short of providing a precise explanation for the fading of the first two extrema during the transition from the quasi--uniform to the filamentary mode in our experiments. Therefore, we consider fast gas heating(FGH), a non-negligible factor that causes discharge instability in high repetition frequency pulse discharge. FGH initiates gas expansion, altering neutral density, elevating the reduced electric field (E/N), and changing reaction rates~\cite{zhao2020volume,popov2021repetitively,zhu2018fast}. Existing research~\cite{leonov2016dynamics} has delved into air disturbances resulting from fast gas heating, revealing the appearance of various shock waves between microseconds and hundreds of microseconds after the discharge pulse. These waves introduce distinct disturbances, showcasing a clear correlation between discharge patterns and shock wave amplitudes.

To comprehend the discharge transition process, the secondary SIWs explain current extrema variations within a single pulse, but they can not fully account for the discharge transition in repetitive frequency discharges. Higher repetition frequencies, with shorter pulse intervals akin to the time scale of airflow disturbance, accumulate the memory effects. Here, the cumulative impact of gas heating emerges as a likely driver for the transition in repeated frequency pulses. A comprehensive investigation is crucial to unravel this intriguing phenomenon.

\begin{figure}[H]
\renewcommand{\figurename}{Fig.}
\centering
\includegraphics[width=83mm]{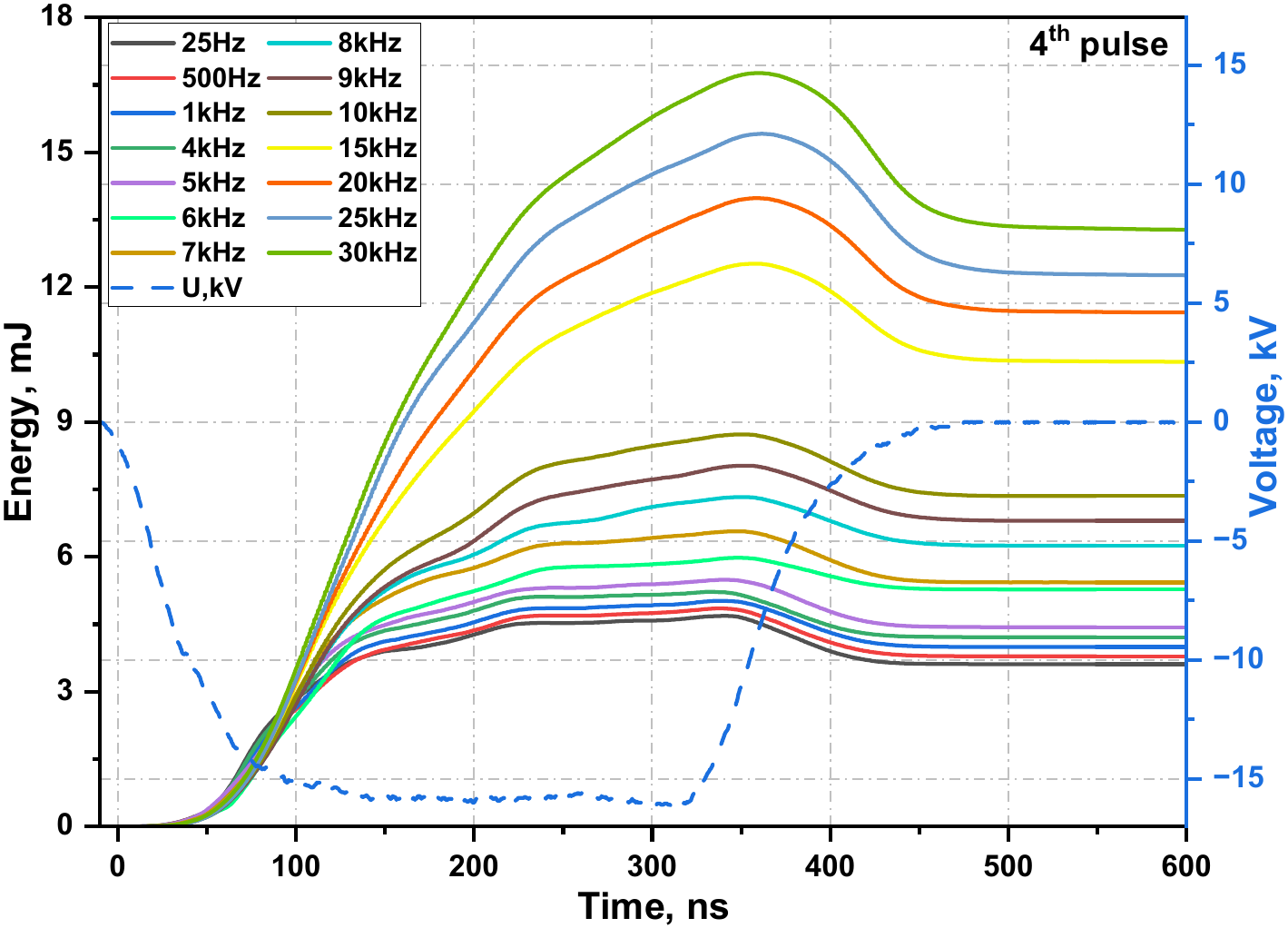}
\caption{Trend of single--pulse deposited energy at different frequencies(25~Hz to 30~kHz) with fixed pulse numbers and voltage amplitude. (The fourth pulse and --16~kV) }
\label{fig8:UIE}
\end{figure}

% introduction for the deposited energy
Another electrical characteristic is the deposited energy. The deposited energy was calculated by integrating the voltage--current product during each single pulse. The correlation between the deposited energy and repetition frequency is illustrated in Fig.~\ref{fig8:UIE}.

% each single line
For a comprehensive overview, the deposited energy consistently rose with the repetition frequency. Considering that deposited energy serves as a metric for discharge intensity, the frequency exhibited a modest promotional effect on the discharge under these conditions. The growth rate underwent a rapid ascent from quasi--uniform to filamentary mode. Notably, at 30~kHz, the maximum deposited energy soared to approximately 13~mJ, marking a threefold surge compared to 25~kHz (3.7~mJ). Within the range of 25~Hz to 5~kHz, the energy deposited per pulse remained relatively lower. However, the transition from 6~kHz to 30~kHz witnessed a significant upswing in deposited energy, indicating a more pronounced promotional effect.

In conclusion, increasing the repetition frequency consistently enhances the deposited energy, especially accentuating the promotional impact on discharge at higher frequencies.

% Effect of Pulse Number
% - deposited energy
% - current 
\subsubsection{Effect of Pulse Number}

% effect of the pulse number (1kHz, 30kHz)
Boosting the repetition frequency enhanced the discharge with a constant number of pulses. Yet, under the fixed frequency, the discharge may not always enhanced with the number of pulses. 

\begin{figure}[H]
\renewcommand{\figurename}{Fig.}
\centering
    \includegraphics[width=83mm]{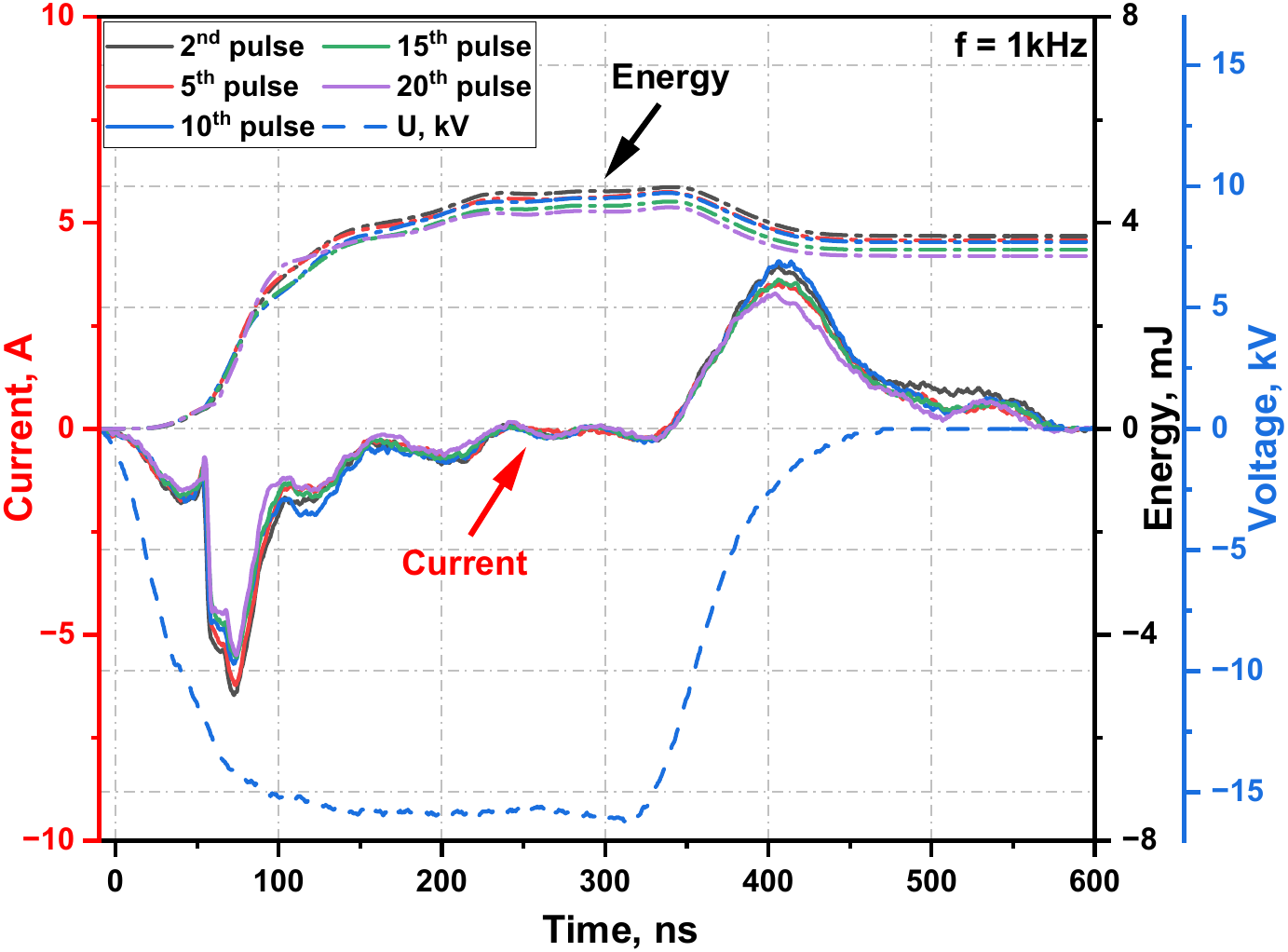}
    \caption{Trend of the current and the deposited energy across a varying number of pulses(2 to 20) with the fixed voltage amplitude(--16~kV). Under this condition, the discharge mode was quasi--uniform discharge. The chain line represents the deposited energy, and the solid line represents the current.}
    \label{fig9:EI1kHz}
\end{figure}

\begin{figure}[H]
\renewcommand{\figurename}{Fig.}
\centering
    \includegraphics[width=83mm]{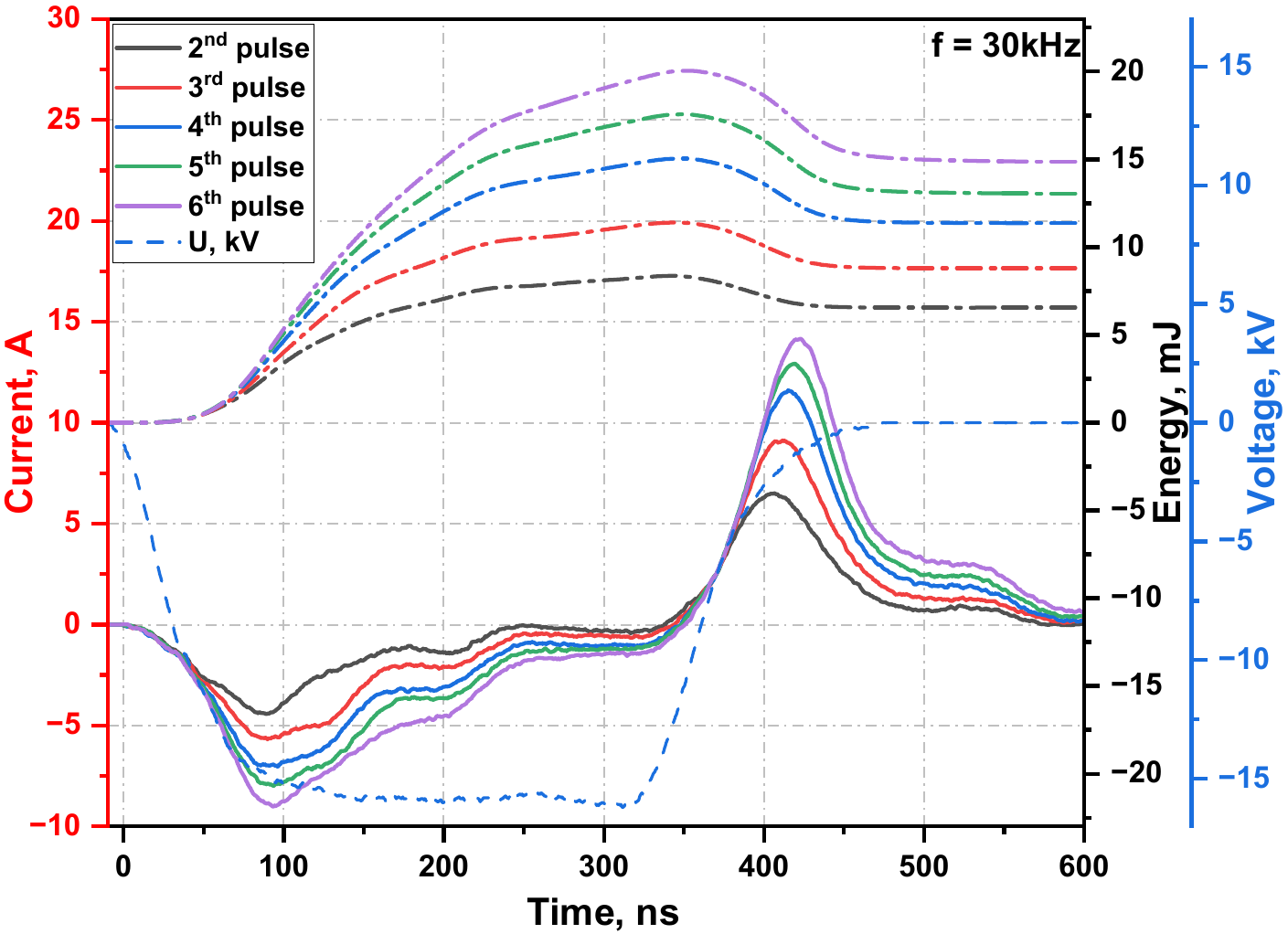}
    \caption{Trend of current and deposited energy across varying the number of pulses(2 to 6) with a fixed voltage amplitude(--16~kV). Under this condition, the discharge mode changed from the transitional to the filamentary mode.}
    \label{fig10:EI30kHz}
\end{figure}

Fig.~\ref{fig9:EI1kHz} and Fig.~\ref{fig10:EI30kHz} illustrated the evolution of the deposited energy varying from the number of pulses at 1~kHz (quasi--uniform) and 30~kHz (filamentary). 

% effect of the pulse number(every single pulse), 1kHz, quasi--uniform
At a consistent repetition frequency, we traced the evolution of the current and the deposited energy across a range of pulses (2 to 20) at 1~kHz, as depicted in Fig.~\ref{fig9:EI1kHz}.

At 1~kHz, With an increase in pulses from 2 to 20, an overlap in the current waveforms and the deposited energy is observed. The primary current peak amplitude was --5.7~A and the secondary one was 3.5~A, showing minimal sensitivity to the number of pulses. Initially, the deposited energy climbed to 4.7~mJ, dipped to 3.6~mJ, and stabilized. Notably, both the primary current peak and stable deposited energy exhibited a slight, frequency--dependent reduction of approximately 6~\%.

The insights from Fig.~\ref{fig9:EI1kHz} suggest a predominant role of the inhibition effect within the pulse interval. As the number of pulses rose, both the current and the deposited energy exhibited a noteworthy decline.

% effect of the pulse number(every single pulse), 30kHz, transitional to filamentary
The effect of the frequency at 30 kHz yielded distinct outcomes. Fig.~\ref{fig10:EI30kHz} illustrates the heightened impact of residual memory factors on subsequent discharges.

% overall view
From a holistic perspective, both the current and the deposited energy exhibited a direct proportionality to the number of pulses. With an increase in pulses from 2 to 6, the stable deposited energy rises from 6.55~mJ to 14.89~mJ, and a noteworthy observation is the deposited energy for a single pulse up to approximately 20~mJ. The discharge inception time of the primary discharge remained constant, while delayed in the secondary discharge, stretching from 338~ns to 351~ns. This delay can be attributed to the brief 50~$\mu$s pulse interval, where a single pre--sequence pulse accumulates additional promotion factors, enhancing subsequent discharges. Consequently, the current takes a longer time to reach zero, leading to a delayed secondary discharge inception time.

\begin{figure}[H]
\renewcommand{\figurename}{Fig.}
\centering
    \begin{subfigure}{81mm}
    \includegraphics[width=81mm]{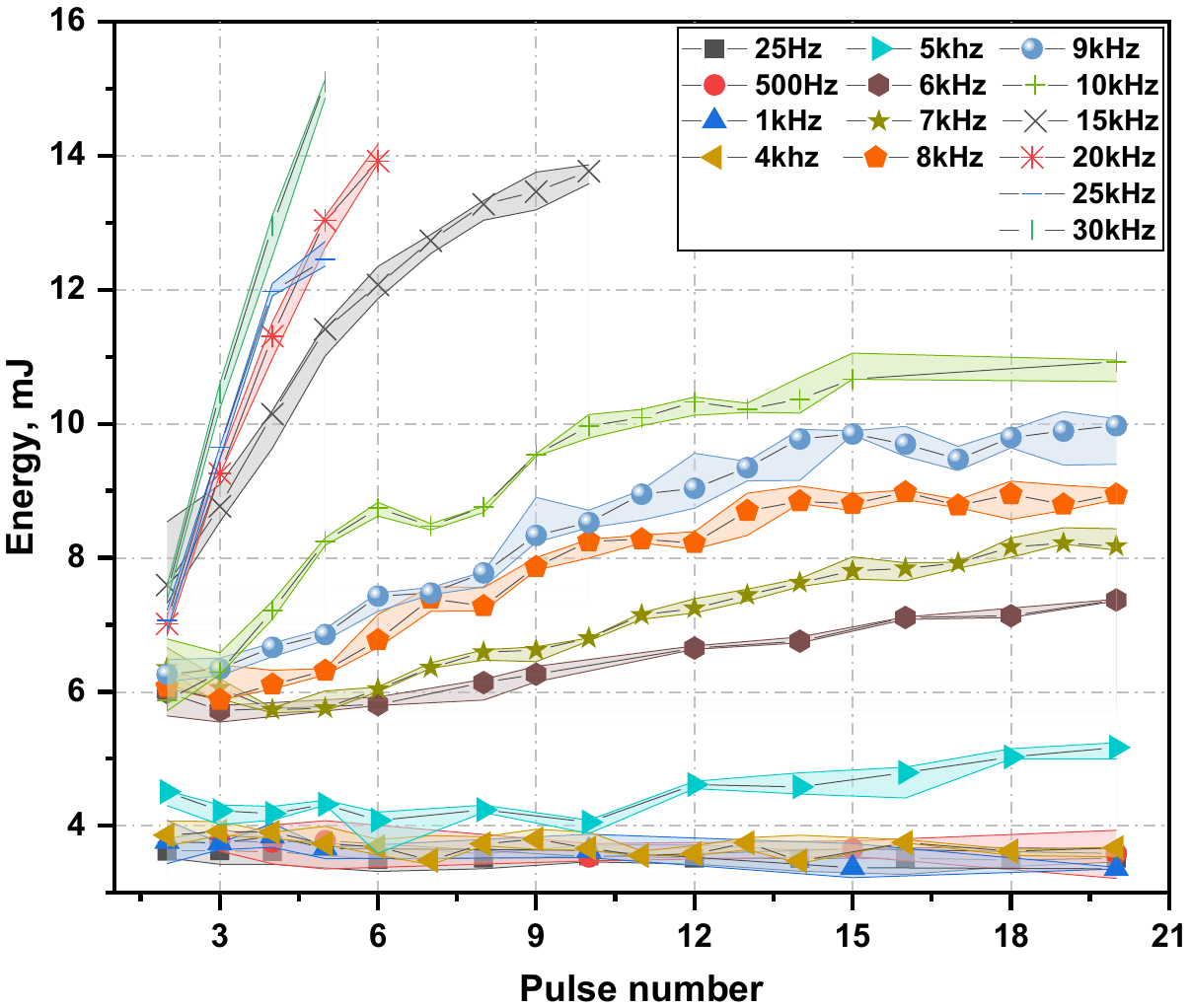}
    \caption{25~Hz to 30~kHz}
    \label{fig11_a}
    \end{subfigure}
    
    \begin{subfigure}{81mm}
    \includegraphics[width=81mm]{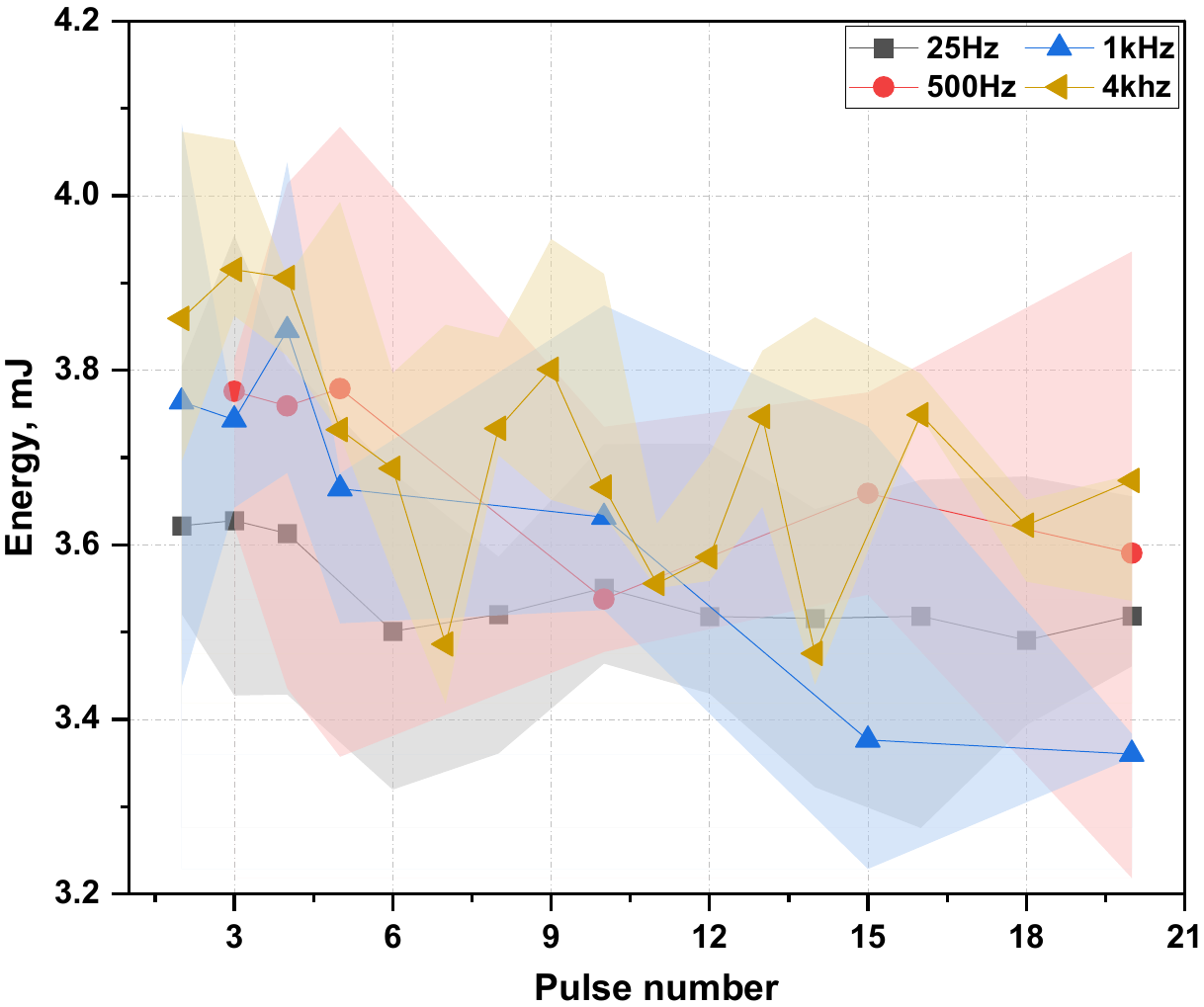}
    \caption{25~Hz to 4~kHz}
    \label{fig11_b}
    \end{subfigure}
\caption{Trend of deposited energy across varying pulse numbers(2 to 20) and repetition frequencies(25~Hz to 30~kHz) with fixed voltage amplitude(--16~kV)}
\label{fig11:Ef}
\end{figure}

% deposited energy changing with the frequency and the pulse number
In exploring the impact of varying pulse parameters on electrical characteristics, we focused on altering a single parameter while keeping others constant. Fig.~\ref{fig7:Current} to \ref{fig10:EI30kHz} depict the distinctive change patterns observed when adjusting either repetition frequency or the number of pulses individually.Intriguingly, simultaneous changes in both pulse parameters yield compelling results, as illustrated in Fig.~\ref{fig11:Ef}.

% overall view
In Fig.~\ref{fig11:Ef}(a), we highlighted three trends aligning with the discharge modes as depicted in Fig.\ref{fig4:distance}. 

1) 25~Hz to 4~kHz (Quasi--uniform mode): The consistently low deposited energy, hovering around 3.7~mJ, remained unaffected by changes in the number of pulses; A closer observation at the 25~Hz to 4~kHz range in Fig.\ref{fig11:Ef}(b) reveals a subtle downward trend of approximately 0.3~mJ; This hints at a noteworthy inhibition effect during longer pulse intervals, emphasizing the persistence of inhibition factors.

% special phenomena

2) 5~kHz to 8~kHz (Transitional mode): In this frequency range, the deposited energy witnessed a noteworthy surge, escalating from 28.8~\% to 52.2~\%. This upswing stems from the heightened pulse frequency, translating to shorter interval time. This fosters the gradual accumulation of promotion factors, offsetting the inhibitory effects. The cumulative impact manifests as a substantial rise in the deposited energy.

Intriguingly, we observed a phenomenon akin to what was discussed in Section 3.1, where deposited energy exhibited a non--monotonic pattern --- initially decreasing and then increasing with the number of pulses. The required pulses for the minimum deposited energy decreased with the frequency. The mechanism behind this phenomenon remains unclear. Significantly, this behavior is intimately linked to the discharge mode transition, manifesting exclusively during transitional modes, while the quasi--uniform and filamentary modes demonstrated monotonic changes.

3) 9~kHz to 30~kHz (Filamentary mode): deposited energy exhibited a monotonically increasing trend, from 59.1~\% to 100.1~\%. Beyond 15~kHz, the flashover occurred as the pulse number continued to increase. At 30~kHz, the stable deposited energy peaked at 15~mJ over five pulses. Promotion factors led with the high repetition frequency (short pulse interval time). The cumulative effect of these factors in each interval propelled the discharge continuously to intensify.

In summary, by adjusting both the repetition frequency and the number of pulses, we observed distinct patterns in the lines, revealing three discernible trends across different frequency ranges. This outcome aligns with the conclusions illustrated in Fig.~\ref{fig4:distance}.

Our exploration into pulse parameters revealed a pivotal role in steering the transition or stabilization of discharge modes. The dynamic interplay of repetition frequency and number of pulses emerged as key influencers.

\subsubsection{The Transition Curve of Discharge Modes}

% Why do we give the transition curve
By investigating the effect of the pulse parameters, we already knew that modulating the pulse parameters can lead to the discharge mode transition.

\begin{figure}[H]
\renewcommand{\figurename}{Fig.}
\centering
\includegraphics[width=83mm]{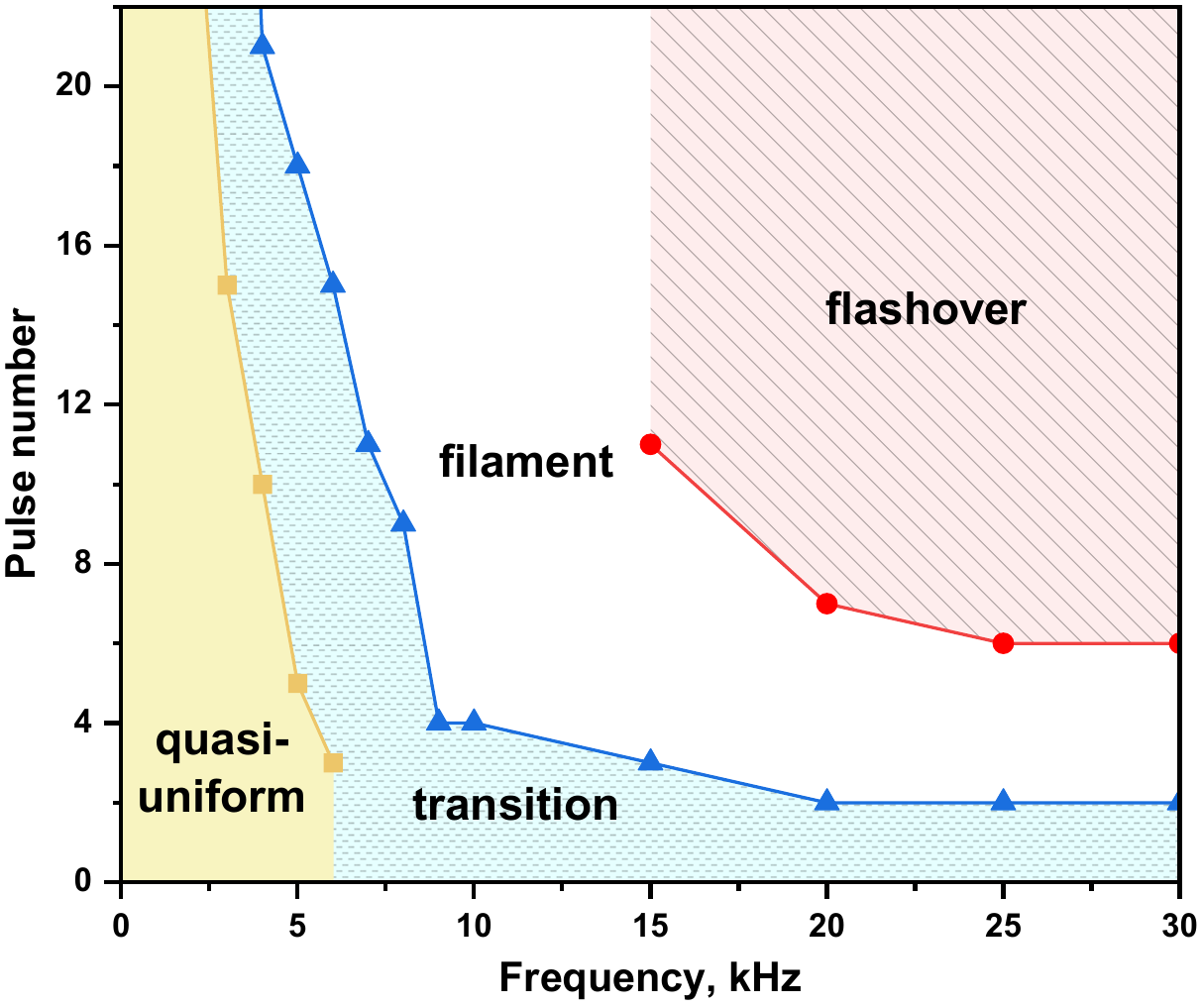}
\caption{The N--F transition curve of discharge modes(quasi--uniform dischagre, transition mode, filamentary discharge, flashover).}
\label{fig12:Transition}
\end{figure}

Fig.~\ref{fig12:Transition} illustrates the N--F transition curve for the discharge modes.

The figure illustrates four distinct discharge modes, namely the quasi--uniform, the transitional, the filamentary, and the surface flashover. It is important to highlight that our study did not include the surface flashover. We should avoid experimenting in the flashover region shown in the figure. The transition, occurring under varying repetition frequencies and pulse numbers, is notably absent below 5~kHz. As the frequency escalates from 6~kHz to 30~kHz, the pulses needed for transition decrease from 21 to 3. Beyond 15~kHz, merely eleven pulses can induce surface flashover. In practical applications, consider insulating the ground electrodes to avert surface flashover.

The quasi--uniform discharge uniformly generates electrons, ions, and active particles, making it suitable for large--scale sterilization, disinfection, and combustion. The filamentary discharge, characterized by high energy density in each channel, is ideal for meeting multi--channel ignition requirements. Precise control allows for versatile applications.

% 4. Conclusion
\section{Conclusion}
This work explored the electrical characteristics, the morphology, and the transition mechanism from quasi–-uniform to filamentary discharge under negative repetitive pulses. The discharge modes were precisely controlled by adjusting repetition frequency and pulse number.

%The memory effect, influenced by surface charge and gas heating disturbance, significantly impacted discharge behavior. Surface charges exhibited time–-dependent inhibitory or promotional effects, while gas heating disturbance introduced instability, disrupting discharge uniformity. These memory effects could weaken or enhance the emission intensity, current, and deposited energy with repetition frequency and pulse number. Higher repetition frequencies required fewer number of pulses for transition. 

Increasing the repetition frequency from 25~Hz to 30~kHz reveals three distinct currents, deposited energy, and propagation trends, corresponding to unique discharge modes: 1) Quasi--uniform mode (25~Hz to 4~kHz): With an inhibitory effect, this mode shows reduced negative surface charges near the HV electrode, featuring three current extrema during the primary discharge. Surface charge saturation is not achieved in this range; 2) Transitional mode (5~kHz to 8~kHz): Initially inhibited, then transforms into a discharge--promoting state. The third current extremum steadily increases, surpassing the first two. Early pulses saturate, but subsequent increases in pulse number disrupt discharge stability without contributing to further surface charge accumulation; 3) Filamentary mode (9~kHz to 30~kHz): This phase exhibits discharge promotion with an increased repetition frequency and pulse numbers. The first two current extrema fade, leaving only the third to increase with frequency.

% key findings
In the transition process, two key findings emerge. Firstly, the current and deposited energy both exhibit a non-monotonic shift exclusively, distinguishing it from the quasi--uniform and the filamentary modes. Secondly, a noteworthy alteration occurs in the count of current extrema during the primary discharge process, transitioning from three to one. This change is specific in the discharge transition process, with no variation in the number of current extrema observed in a stable single discharge mode, only changes in amplitude. This intriguing phenomenon may be attributed to the influence of secondary SIWs, establishing a crucial link between these findings and the discharge transition. 

Moreover, the N--F curve of the discharge transition was given. As illustrated in the figure, higher frequencies and pulse numbers correlate with a more pronounced shift toward filamentous discharge. This observation holds significant implications for practical applications as a foundation for controlling the pulse parameters.

Conducting meticulous qualitative and quantitative experiments across a broad frequency range, the intricate discharge morphology and electrical characteristics are revealed. This study enhances our insights into diverse discharge modes, unveiling unique phenomena. The resulting transition curve distinctly outlines pulse parameter groups associated with the specific discharge modes. This curve aids in optimizing pulse power supply and control schemes in practical applications.

\section*{Acknowledgments}
The research was supported by the Fundamental Research Funds for the Central Universities, China (Grant No. xtr052023003) and the State Key Laboratory of Electrical Insulation and Power Equipment, China (No. EIPE23114).
\section*{References}

\bibliography{surface_discharge_transition}

\end{document}